# A unique focusing property of a parabolic mirror for neutrons in the gravitational field: geometric proof

S. Masalovich

Technische Universität München, Forschungsneutronenquelle Heinz Maier-Leibnitz (FRM II),
Lichtenbergstr. 1, D-85747 Garching, Germany
E-mail: Sergey.Masalovich@frm2.tum.de
Submitted to arXiv, April 2014

**Abstract**

An extraordinary focusing property of a parabolic mirror for ultracold neutrons in the presence of the gravitational field was first reported by A. Steyerl and co-authors. It was shown that all neutrons emitted from the focus of the mirror will be reflected back upon the same focus point passing, in between, a point of return in the gravitational field. The present note offers a complementary geometric proof of this feature and discusses some implications.

Keywords: Ultracold neutrons, gravitational field, parabolic mirror, neutron microscope.

**Introduction**

In 1985, the group of scientists led by A. Steyerl at the Technische Universität München reported successful operation of a unique neutron microscope using ultracold neutrons (UCN) [1]. The ability of such neutrons to reflect specularly from surfaces of many materials at any angle of incidence makes it possible, in principle, to use common optical schemes for designing a neutron reflecting microscope. However, unlike common optics, in UCN optics neutron trajectories are curved by the Earth's gravity. The curvature of every trajectory (flight parabola) depends on the neutron velocity and the initial angle at an object thus resulting in appearance of considerable chromatic aberrations.

With the aim to correct these aberrations the authors developed the very special two-mirror optical system with a common vertical axis. In this system neutrons first pass through an object and move downwards, then reflect upwards from a concave parabolic mirror, pass the highest points of flight parabolas and, on their way down, reflect from a convex spherical mirror. The achromatism of that system is mainly resulted from the unique property of the focal point of a parabolic mirror discovered by the authors. In the first paper [1] this property was reported for paraxial rays, but in the next paper [2] it was already stated: "A particle emanating from the focal point in any direction and at any given speed is reflected back into its origin by the mirror which we assume to be infinite extent". In addition, the authors reported "the unique property of constant flight-time between the coincident object and image, irrespective of initial flight direction". The both properties were confirmed by extensive algebraic calculations and computer simulations.

In this note, we present a geometric proof of the extraordinary property of a parabolic mirror mentioned above. Although the proof below will be given for ultracold neutrons in the Earth's gravitational field, it remains true for any massive particles in a homogeneous force field (e.g., an electron or ion in an external electric field).

**Geometric proof**

First we prove that the focus of a parabolic mirror will be imaged back upon itself by that mirror. Because of the axial symmetry we only consider trajectories in the plane through the axis of the parabolic mirror. The equation for the mirror's surface in the frame of reference, in which the origin is located at the focus, is given by

$$z = \frac{x^2}{4f} - f, \tag{1}$$

where $f$ is the focal length of the parabolic mirror. The parabola opens up and the optical axis is arranged along the vertical axis Z, whereas the axis X is arranged horizontally. Let us now write the equation of a neutron trajectory for the neutron emitted from the origin with the velocity $v$ and angle $\alpha$ with the vertical axis:

$$\begin{aligned} x(t) &= v \cdot \sin\alpha \cdot t \\ z(t) &= v \cdot \cos\alpha \cdot t - \frac{gt^2}{2} \end{aligned} \tag{2}$$

Here $g$ is the acceleration due to gravity and $t$ is a flight time. From Eq. (2) one can get

$$z_1 = \frac{2g}{4v^2 \sin^2\alpha} \cdot x_1^2 - \frac{v^2 \sin^2\alpha}{2g}, \tag{3}$$

where we used the substitutions

$$\begin{aligned} x_1 &= x - \frac{v^2 \sin 2\alpha}{2g} \\ z_1 &= -z + \frac{v^2 \cos 2\alpha}{2g} \end{aligned} \tag{4}$$

By comparing Eqs. (1) and (3), one can see that the neutron trajectory is a parabola with the focal length equal to $v^2 \sin^2\alpha/2g$. Additionally, from Eq. (4) we find the coordinates of the focus of that flight parabola: ($v^2 \sin 2\alpha/2g$, $v^2 \cos 2\alpha/2g$). Thus we arrive to the conclusion (well-known, for example, in electron optics [3] for electrons in a homogeneous electric field) that focuses of all parabolic trajectories of particles emitted from the origin with the velocity $v$ lie on a circle of radius $R = v^2/2g$ with a center in the origin.

We will now examine imaging properties of a parabolic mirror for neutrons in the Earth's gravitational field. Fig.1 shows the neutron trajectory (line I) passing through the focus O of the parabolic mirror and the trajectory of the same neutron (line II) after being reflected specularly from the mirror at the point A.

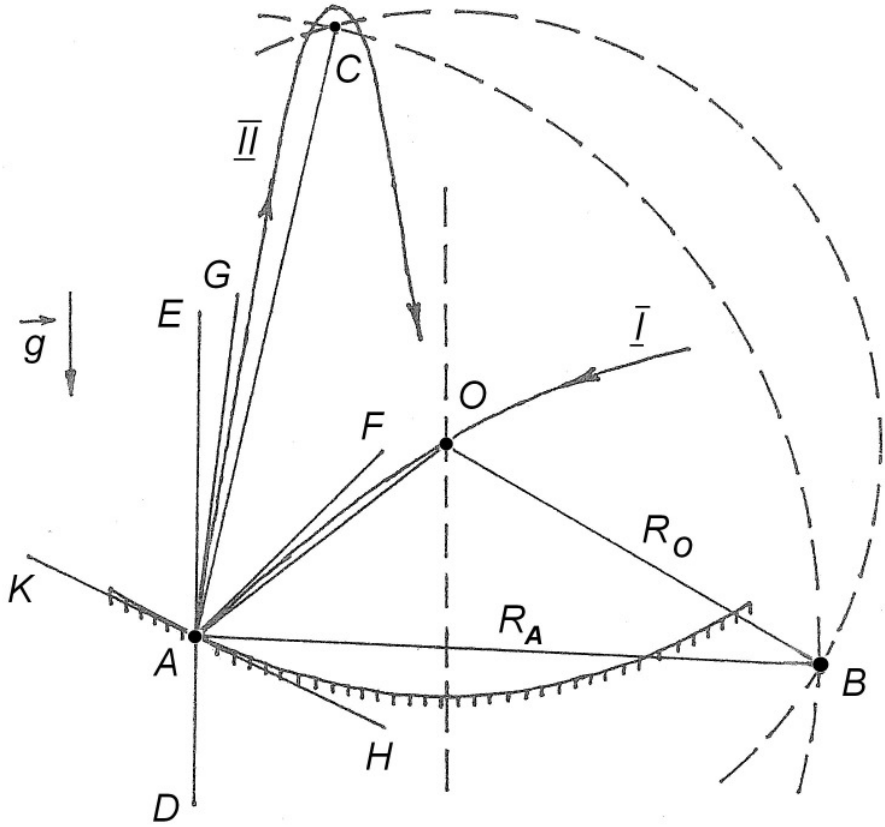

Fig.1 Reflection of a neutron by a parabolic mirror

Let us denote by $v_0$ the neutron velocity at the point O. Then, as it follows from the above results, the focus B of the parabola I lies on the circle of radius $R_0 = v_0^2/2g$ with the center at O. Since the parabolas I and II pass through the common point A, and at this point they both are characterized by the velocity $v_A$, then the focuses of these two parabolas lie on the same circle of radius $R_A = v_A^2/2g$ with the center at A. The focus of the line II is marked by the letter C in Fig.1. We are to show that C lies on the circle $R_0$. Let us draw the vertical line DE through the point A, which is parallel to the axes of the parabolas I and II, and parabolic mirror. Besides, through this point we draw tangents to the curve I (AF), to the curve II (AG), and to the parabolic mirror (HK). Due to specular reflection one can write

$$\angle KAG = \angle HAF .\tag{5}$$

By the properties of a parabola, we have

$$\angle KAE = \angle DAH = \angle HAO ,\tag{6}$$

$$\angle BAF = FAE ,\tag{7}$$

$$\angle GAE = GAC .\tag{8}$$

From Eqs. (5-6) it follows that

$$\angle GAE = \angle FAO .  \qquad (9)$$

Rewriting Eq. (7) in terms of the results (8) and (9), we obtain

$$\angle CAO = \angle BAO .  \qquad (10)$$

Hence we have shown that the points B and C not only lie on the same circle with the center at A, but also they are symmetric about the line OA. Since the point B belongs to the circle with the center at O, then the point C belongs to the same circle and $OC = v_0^2/2g$ as illustrated in Fig.1. We now prove that the line II passes through O. Indeed, by the energy conservation law, the neutron velocity equals $v_0$ at the point of intersection of a trajectory with the horizontal line passing through O. Assuming now that the neutron trajectory crosses this horizontal line not at O but at some neighboring point $O_1$, we arrive to the conclusion that $O_1C$ has to be equal to $v_0^2/2g$. Since we have already shown that $OC = v_0^2/2g$, then it follows that $O_1$ coincides with O and the reflected parabola II passes through the focus of the mirror. If we now recall that as initial conditions we took an arbitrary neutron velocity and angle with the vertical axis at O, we can conclude that the proved statement is true for any neutron trajectory.

Yet another important conclusion can be derived from the above result. If a neutron does not pass through the focus of a parabolic mirror at the beginning of its movement, then it will never pass through this point in the course of multiple successive reflections from the mirror. We will show in our next paper [4] that the area, accessible to a neutron trajectory in this instance, is bounded by the surface of the parabolic mirror and by two other parabolic surfaces (envelopes) with all three focuses lying at the same point. The parameters of the envelopes depend on initial coordinates and velocity vector of a given neutron. As a result, two different situations may take place: Either both parabolas (envelopes) open down, or one parabola opens down while other opens up. This means that if a neutron trajectory at the beginning intersects the axis of a parabolic mirror above (under) the focus, then it will always intersect this axis only above (under) the focus in the process of multiple reflections from the mirror surface. It might be well to notice that in either case there will be an area around the common focus that is surrounded by neutrons but not accessible to them.

**Flight time along a closed trajectory**

Now we will show that the time of propagation of a neutron along the closed trajectory discussed above (e.g., line I + line II) does not depend on an initial angle. Let $v_0$ be the neutron velocity at O. From the parametric formula of a flight-parabola (see Eq. (2)) we obtain

$$x^2 + \left(z + \frac{gt^2}{2}\right)^2 = (v_0 t)^2 \qquad (11)$$

This equation does not include the angle that the neutron velocity vector makes with the vertical axis at O. Furthermore, one can see that for every given *t* all neutrons, emitted simultaneously from O in all directions, reach a circle (a sphere in 3D space) of radius $R = v_0 t$ and with a center at the point $(0, -gt^2/2)$. The point in time *t*, at which a neutron trajectory crosses the parabolic mirror, can be found by solving simultaneously Eqs. (1) and (11). This leads to the equation

$$\left(\frac{g}{2}\right)^2 t^4 - \left(v_0^2 - gz\right)\cdot t^2 + (z+2f)^2 = 0 \qquad (12)$$

It has been shown above that there are two parabolic neutron trajectories (e.g., lines I and II in Fig.1) that pass through the point O and cross the mirror surface at a given point. Therefore Eq. (12) has to have two physically admissible solutions: $t_1$ and $t_2$. The sum of these two values is just a time of propagation along a closed trajectory. By the properties of quadratic equations, we have

$$t_1^2 + t_2^2 = \left(\frac{2}{g}\right)^2 \left(v_0^2 - gz\right)$$
$$t_1^2 \cdot t_2^2 = \left(\frac{2}{g}\right)^2 (z+2f)^2 \qquad (13)$$

From this immediately follows the final result

$$t = t_1 + t_2 = \frac{2}{g}\sqrt{v_0^2 + 2gf} \qquad (14)$$

We see that the time of propagation does not depend on the coordinates of the point where a neutron trajectory crosses the mirror. Thus it does not depend on the initial angle of the trajectory, which proves the statement.

In the earlier paper [5] we studied the linkage between achromatism of a neutron imaging system and the property of trajectories to be of equal propagation time in such system. In that paper the geometric proof (see above) of the property of a parabolic mirror was first presented. For a more common case when an optical system has a size $l$ such that $gl/v^2 \ll 1$, it was shown that the condition for an achromatic imaging mirror system for neutrons is that a neutron image coincides with a "light-optics" image in the same system. Such coincidence may serve as a guide for building an achromatic neutron focusing system (e.g., microscope).

**Conclusion**

In the present note the extraordinary properties of a parabolic mirror were established in the case of axial symmetry. Yet it is clear from the proof given above that equivalent properties will be observed for a parabolic cylinder as a mirror surface. The focal axis of such cylinder should lie in a horizontal plane and the cross section of that surface would be a parabola which opens up. Obviously, the component of a neutron velocity that is parallel to the focal axis will not vary as a neutron reflects from the surface. In that case all the proofs remain true in the sense that instead of a real velocity vector we have to use a projection of that vector onto a plane which is normal to the focal axis of the parabolic cylinder. From this it follows that all neutrons emitted from a given point on the focal axis will always cross this axis in their succeeding propagation along such kind of a neutron guide. Consequently, one can think about possible applications of the discussed properties of a parabolic mirror not only for ultracold but also for cold neutrons.

**References**

[1] P. Herrmann, K.-A. Steinhauser, R. Gähler, A. Steyerl, W. Mampe, "Neutron microscope", Physical Review Letters **54** (1985) 1969-1972

[2] A. Steyerl, W. Drexel, T. Ebisawa, E. Gutsmiedl, K.-A. Steinhauser, R. Gähler, W. Mampe, P. Ageron, "Neutron microscopy", Revue Phys. Appl. **23** (1988) 171-180

[3] W. Glaser, "Grundlagen der Elektronenoptik", Springer-Verlag, Wien (1952)

[4] S. Masalovich, "Gravity induced patterns drawn by a neutron trajectory", in preparation

[5] S. Masalovich, "On the "ideal" configuration of a neutron microscope with ultracold neutrons", Preprint Kurchatov Institute of Atomic Energy IAE-5341/14, Moscow (1991)